\documentclass{jpsj-suppl}
\usepackage{txfonts} 

\newcommand{\GeV}{\mathrm{GeV}}
\newcommand{\pipipi}{\pi^-\pi^+\pi^-}
\newcommand{\pipi}{\pi^+\pi^-}
\newcommand{\widthTwo}{.33\columnwidth}

\title{Extraction of the $\pi^+\pi^-$ Subsystem in Diffractively Produced $\pi^-\pi^+\pi^-$ at \textsc{Compass}}

\author{Fabian \textsc{Krinner}$^{1}$ for the \textsc{Compass}-collaboration}

\inst{$^{1}$Technische Universit\"at M\"unchen, Physik-Department, E18, James-Franck-Str. 1, 85748 Garching, Germany}

\email{fabian-krinner@mytum.de}

\recdate{September 26, 2016}

\abst{The \textsc{Compass} experiment at CERN has collected a large data sample of 50 million diffractively produced $\pi^-\pi^+\pi^-$ events using a $190\,$GeV$/c$ negatively charged hadron beam. 
The partial-wave analysis (PWA) of these high-precision data reveals previously unseen details. The PWA, which is currently limited by systematic uncertainties, is based on an isobar model, where multi-particle decays are described as subsequent two-body decays and where a prior-knowledge parametrization for the intermediate two-pion resonances has to be assumed -- usually a Breit-Wigner amplitude -- thus increasing systematic uncertainties, due to the concrete choice of the parametrization. We present a novel method, which allows to extract isobar 
amplitudes directly from the data in a less biased way. The focus lies on the scalar $\pi^+\pi^-$ subsystem, where a previous analysis found a signal for a new axial-vector state 
$a_1(1420)$ decaying into $f_0(980)\pi$.}

\kword{\textsc{Compass}, Partial-Wave Analysis, Isobar Model, $a_1(1420)$}

\begin{document}
\maketitle


\section{Introduction}
\label{sec::compass}
\textsc{Compass} is a two-stage multi-purpose spectrometer, located at CERN's Pr\'evessin site, which employs secondary hadron or tertiary muon beams from the Super Proton Synchrotron. Its large 
acceptance over a wide kinematic range allows \textsc{Compass} to study a broad physics program including, amongst others, light-meson spectroscopy, which is the focus here.

The particular channel of interest is $\pi^-p\to\pi^-\pi^+\pi^-p$, for which \textsc{Compass} collected a data set consisting of approximately $50$ million events. 

\section{Analysis method}
\label{sec::pwa}
\subsection{The Isobar Model}
\label{sec::isobar}
To analyze the process $\pi^-p\to X^-p\to\pi^-\pi^+\pi^-p$ we use the isobar model, which assumes that the appearing intermediate $3\pi$ state $X^-$ does not decay directly into $\pipipi$, 
but undergoes subsequent two-particle decays until it ends up in the final state: $X^-\to\xi^0\pi^-\to\pipipi$. The intermediate two-pion state $\xi^0$ is called the isobar.

\subsection{Conventional PWA}
\label{sec::conventional}
The conventional PWA expands the complex decay amplitude, which describes the measured intensity distribution $\mathcal{I}$, into partial waves \cite{Adolph:2015tqa}:\
\begin{equation}
\label{eq::intens}
\mathcal{I}(\vec\tau;  m_{3\pi}, t^\prime) = \left| \sum_\mathrm{waves} \mathcal{T}_\mathrm{wave}(m_{3\pi}, t^\prime) \Psi_\mathrm{wave}(\vec\tau; m_{3\pi}) \right|^2
.\end{equation}
The production amplitudes $\mathcal{T}_\mathrm{wave}$ depend on the invariant mass $m_{3\pi}$ of the $\pipipi$ system and on the reduced squared four-momentum transfer $t^\prime$. They are fitted to the data
in bins of their kinematic variables using an extended maximum likelihood fit.

For constant $m_{3\pi}$ and $t^\prime$, the decay amplitudes $\Psi_\mathrm{wave}$ depend on $5$ kinematic variables, that define the $3\pi$ kinematics and are represented by $\vec\tau$, while the angular part alone is given by $\vec\Theta$. 
The decay amplitudes are known functions, which have to be put into the analysis model beforehand. They consist of a mass-dependent 
part $\Delta_\xi(m_{\pi^+\pi^-})$ which depends on the mass of the $\pipi$ subsystem, and an angular part $\mathcal{K}(\vec\Theta)$:
\begin{equation}
\Psi_\mathrm{wave}(\vec\tau;m_{3\pi}) = \Delta_\xi(m_{\pipi};m_{3\pi})\mathcal{K}(\vec\Theta;m_{3\pi}) + (\mathrm{Bose\ symm.}).
\end{equation} 
The angular-momentum quantum numbers appearing in a partial wave completely determine the function $\mathcal{K}(\vec\Theta)$. 

The complex function $\Delta_\xi(m_{\pi^+\pi^-};m_{3\pi})$ describes the complex amplitude of the corresponding isobar $\xi$ and usually has to be known without any free parameters. In the simplest cases, single Breit-Wigner amplitudes are used.
Since no unique parametrizations for these amplitudes are given by theory and different models are available, the choice of a particular parametrization introduces a model bias. 

A conventional PWA of this type, which was performed on the data-set collected by the \textsc{Compass} spectrometer, uses a set of $88$ waves \cite{Adolph:2015tqa} .
\subsection{Freed-isobar PWA}
\label{sec::novel}
In order to circumvent this problem we introduce a novel method, which was inspired by Ref. \cite{Aitala:2005yh}. This method allows us to extract isobar amplitudes in bins of $m_{\pi^+\pi^-}$ directly from the data. To this end, the fixed parametrizations are replaced by sets of piece-wise 
constant functions:
\begin{equation}
\Pi_\mathrm{bin}(m_{\pi^+\pi^-}) = \left\{\begin{array}{l} 1 \mathrm{\ if\ }m_{\pi^+\pi^-}\mathrm{\ lies\ in\ the\ corresponding\ mass\ bin,`}\\
							      0 \mathrm{\ otherwise.}\end{array}\right
.\end{equation}
These binned functions replace the fixed isobar amplitudes:
\begin{equation}
\Delta_\xi(m_{\pi^+\pi^-}) \to \sum_\mathrm{bins} \Pi_\mathrm{bin}(m_{\pi^+\pi^-})
.\end{equation}
The set of bins covers the whole kinematically allowed $m_{\pi^+\pi^-}$ mass range. With this replacement, equation (\ref{eq::intens}) reads:
\begin{equation}
 \mathcal{I}(m_{\pi^+\pi^-}, \vec\Theta; m_{3\pi}, t^\prime) = \left| \sum_\mathrm{waves} \sum_\mathrm{bins} \mathcal{T}_\mathrm{wave}(m_{\pi^+\pi^-}^\mathrm{bin};m_{3\pi}, t^\prime)\left[\Pi_\mathrm{bin}(m_{\pi^+\pi^-})\mathcal{K}_\mathrm{wave}(\vec\Theta)\right]+(\mathrm{Bose\ symm.})\right|^2
.\end{equation}
The piece-wise constant isobar amplitudes effectively behave like independent partial waves and their corresponding production amplitudes now also encode information about th $m_{\pi^+\pi^-}$ dependence of the isobar amplitudes. 
Therefore, the same fit procedure as in the conventional approach can be used. We call this new approach {\it freed-isobar} PWA.

A freed-isobar wave is named after the folloing scheme:
\begin{equation}
J^{PC}M^\varepsilon [\pi\pi]_{J^{PC}_\xi} \pi L
,\end{equation}
where $J^{PC}$ are the spin and eigenvalues und parity and generalized charge conjugation of the $3\pi$ system, while $M^\varepsilon$ are its spin projection and reflectivity. The term $[\pi\pi]$ denotes a freed-isobar wave with 
spin and eigenvalues und parity and charge conjugation of $J^{PC}_\xi$. Finally, $L$ is the orbital angular momentum between the isobar and the bachelor $\pi$.

\section{First Application}
\label{sec::results}

The analysis presented in the following employs $3$ freed-isobar waves: $0^{-+}0^+[\pi\pi]_{0^{++}}\pi S$, $1^{++}0^+$ $[\pi\pi]_{0^{++}}\pi P$ and $2^{-+}0^+[\pi\pi]_{0^{++}}\pi D$. Due to quantum numbers 
of the $\pi^+\pi^-$ subsystem, these waves describe seven waves in the conventional scheme. Therefore, the final model consists of $81$ fixed and $3$ freed-isobar waves.

\subsection{$0^{-+}0^+[\pi\pi]_{0^{++}}\pi\,S$ Wave}
\label{sec::zmp}
\begin{figure}[h]
\begin{center}
\includegraphics[width=\widthTwo]{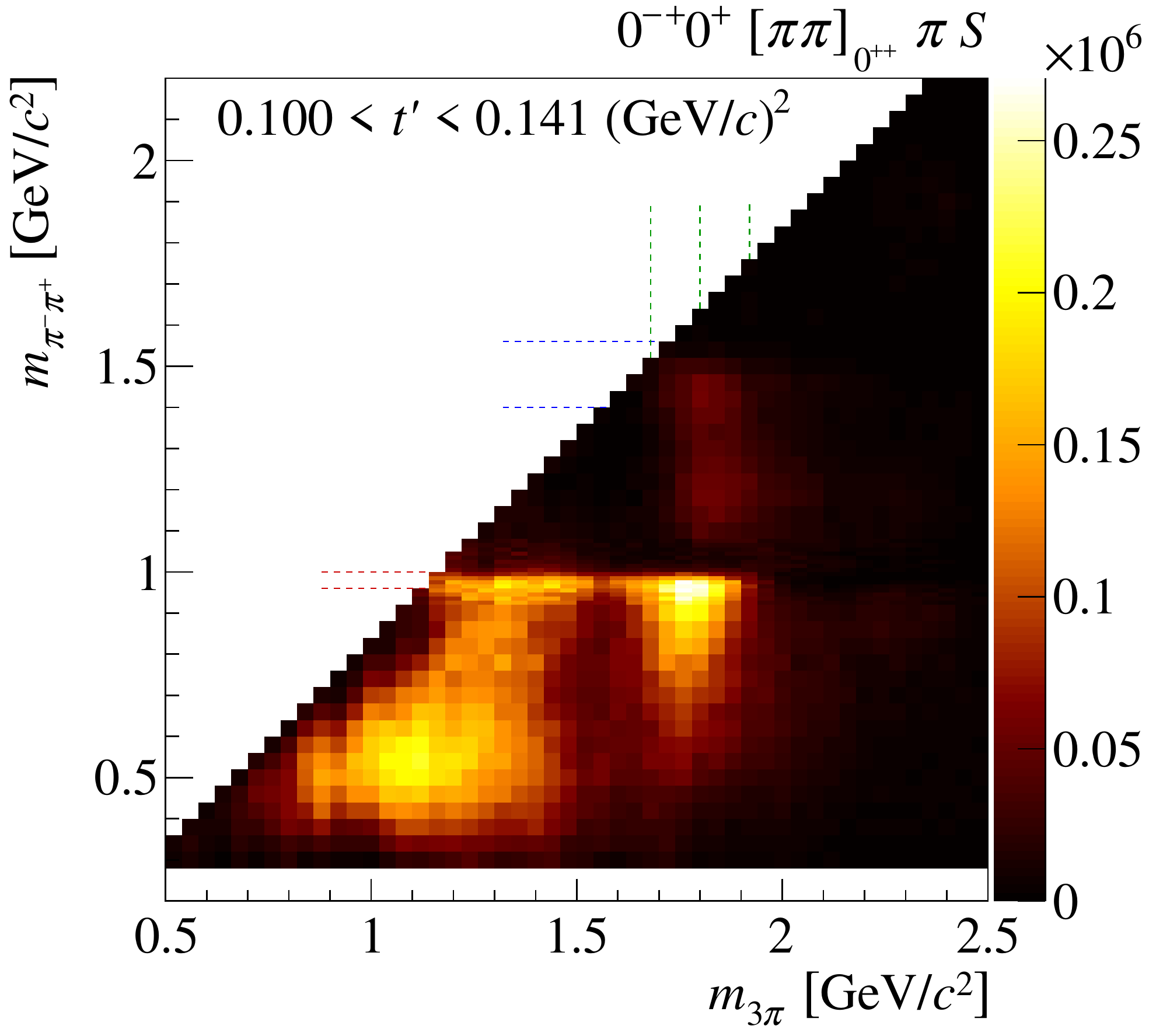}
\includegraphics[width=\widthTwo]{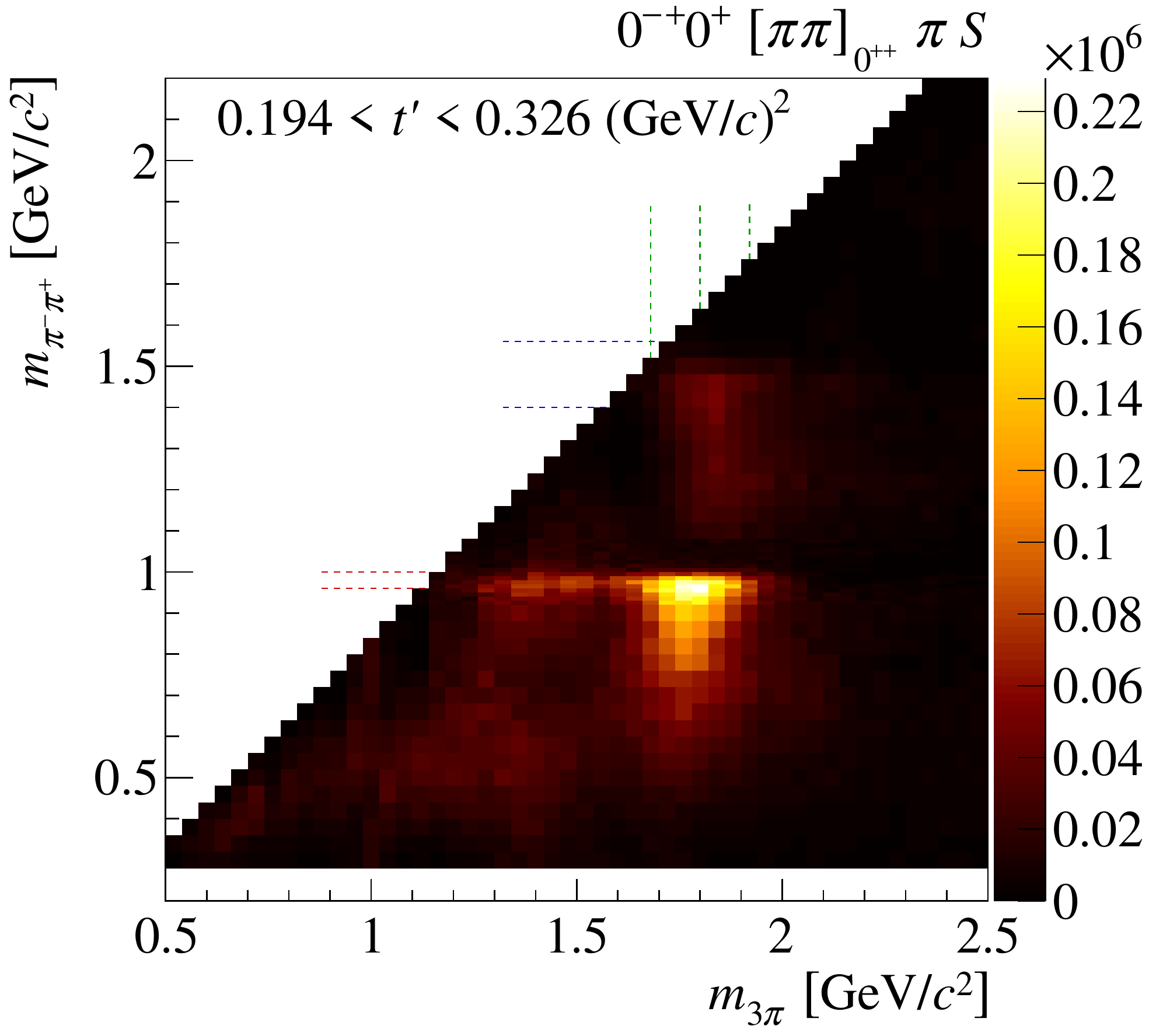}
\end{center}
\caption{Intensity distribution of the $0^{-+}0^+[\pi\pi]_{0^{++}}\pi\,S$ wave as a function of $m_{3\pi}$ and $m_{\pipi}$ for two regions of $t^\prime$ \cite{Adolph:2015tqa}.}
\label{fig::0mp}
\end{figure}

The $0^{-+}0^+[\pi\pi]_{0^{++}}\pi\,S$ wave is able to describe all three isobars that are used in the conventional PWA: the $f_0(500)$, the $f_0(980)$, and the $f_0(1500)$. Fig. \ref{fig::0mp} shows the two-dimensional intensity distribution \\ $\left|\mathcal{T}_\text{wave}(m_{3\pi},m_{\pi^+\pi^-})\right|^2$
for this wave for two bins in $t^\prime$.

The most striking feature is a peak corresponding to the decay $\pi(1800)\to f_0(980)\pi$. A smaller peak corresponding to $\pi(1800)\to f_0(1500)\,\pi^-$ is also visible. Broad structures appear at low $2\pi$ and $3\pi$ masses and low $t^\prime$,
which are probably of mostly non-resonant origin.

\begin{figure}[h]
\includegraphics[width=.33\columnwidth]{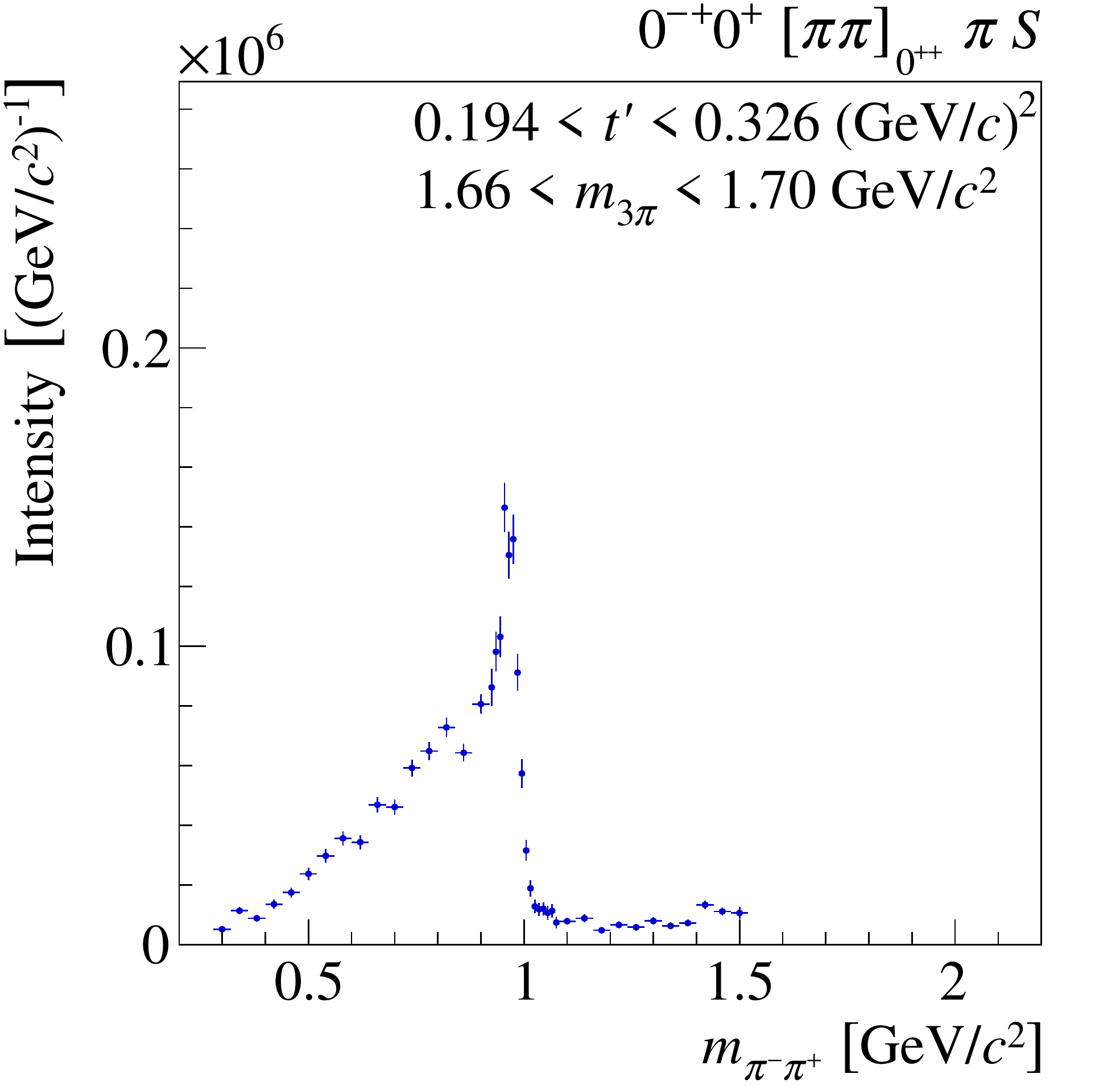}
\includegraphics[width=.33\columnwidth]{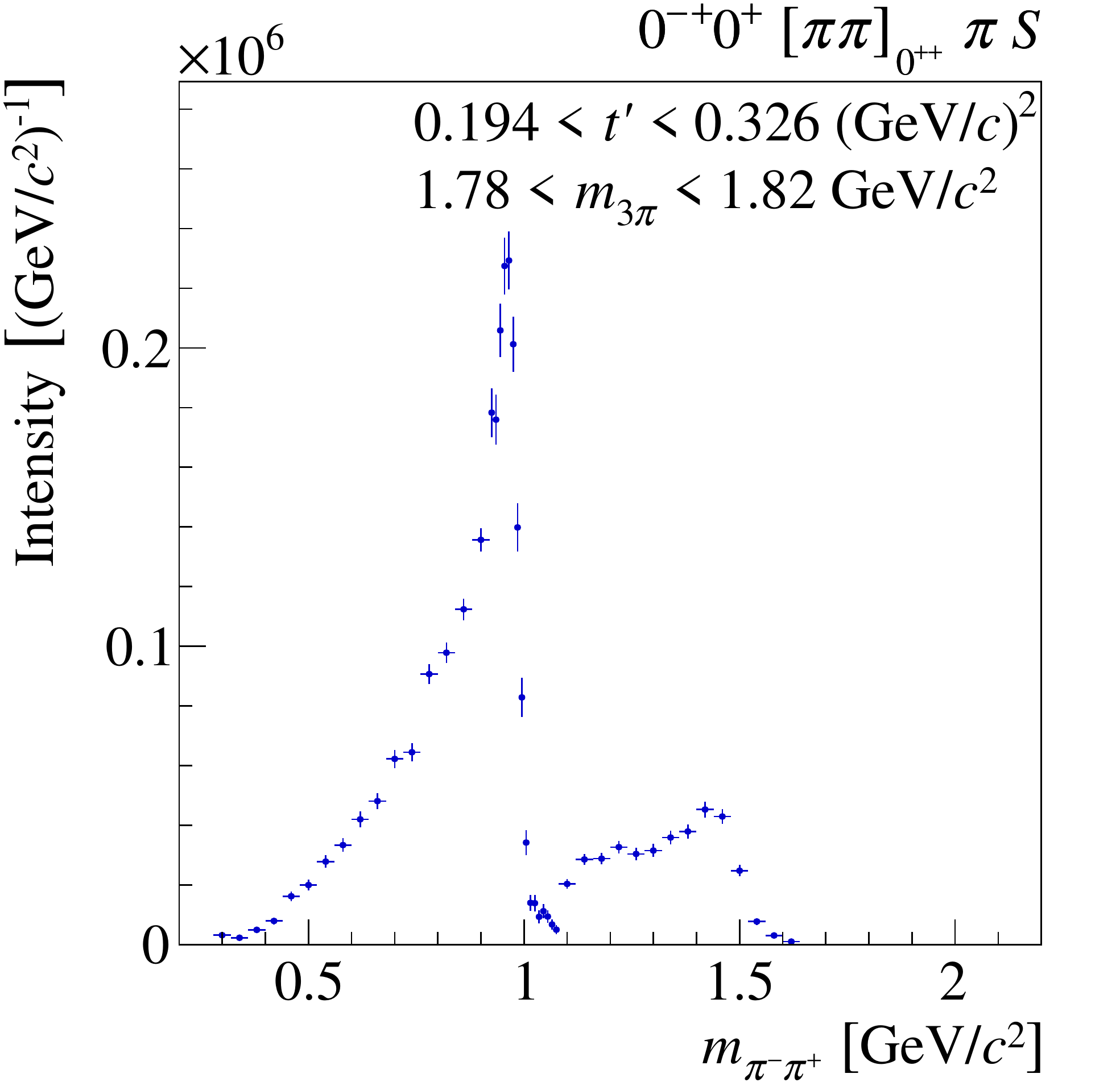}
\includegraphics[width=.33\columnwidth]{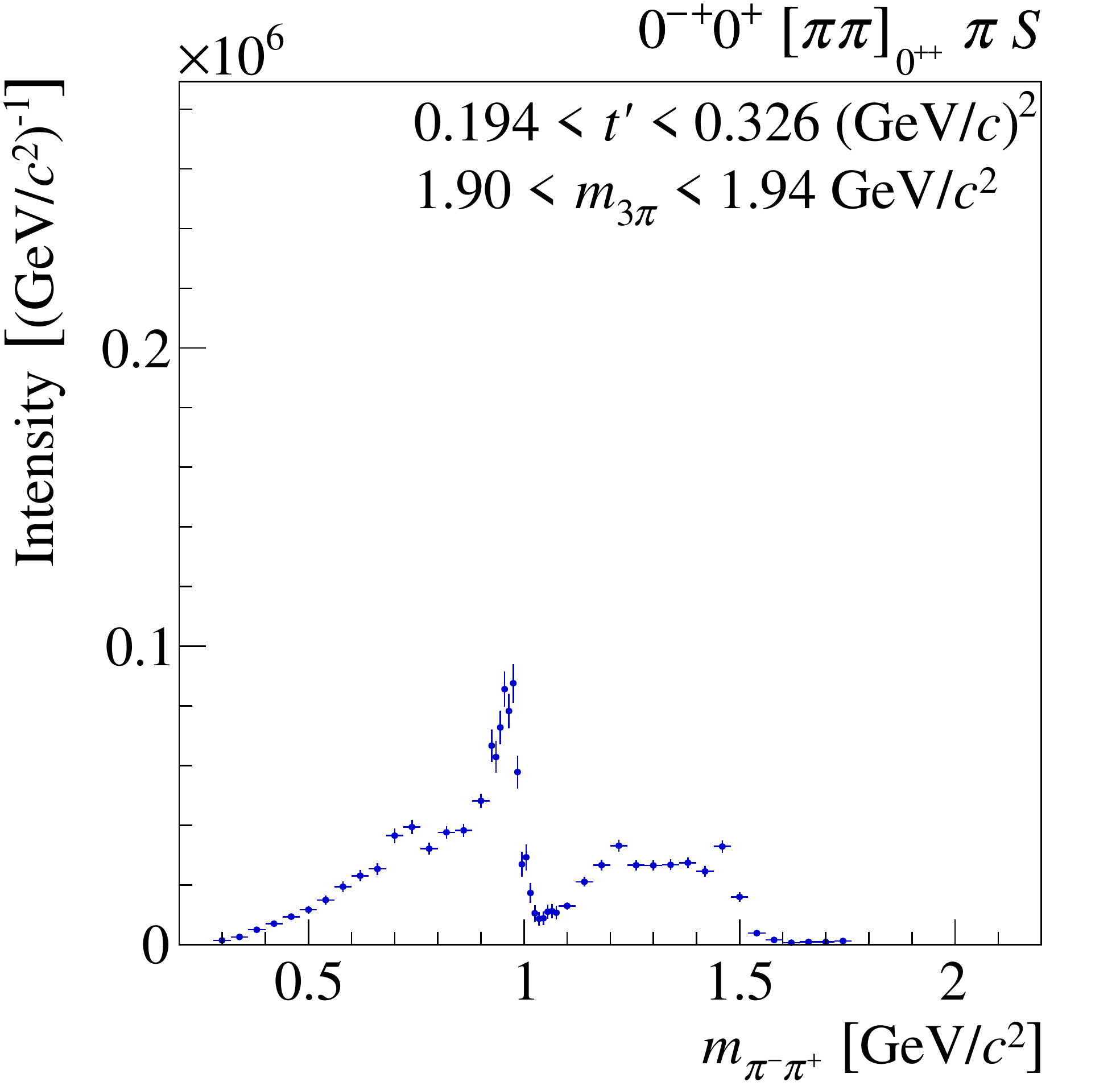}\\
\caption{$0^{-+}0^+[\pi\pi]_{0^{++}}\pi S$ intensity distributions for three bins of $m_{3\pi}$, below, on, and above the $\pi(1800)$ resonance \cite{Adolph:2015tqa}.}
\label{fig::0mpInt}
\end{figure}

\begin{figure}[h]
\includegraphics[width=.33\columnwidth]{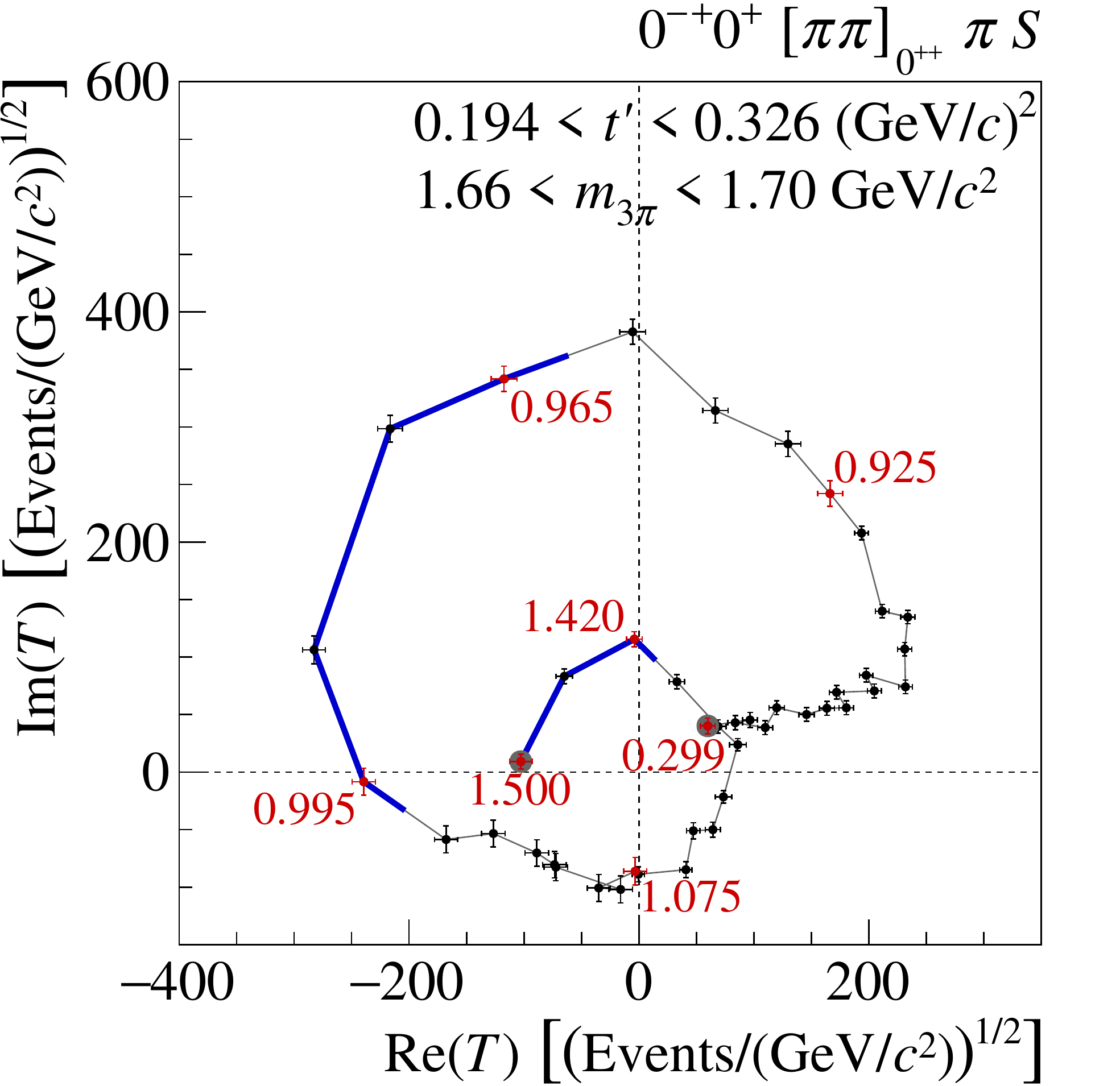}
\includegraphics[width=.33\columnwidth]{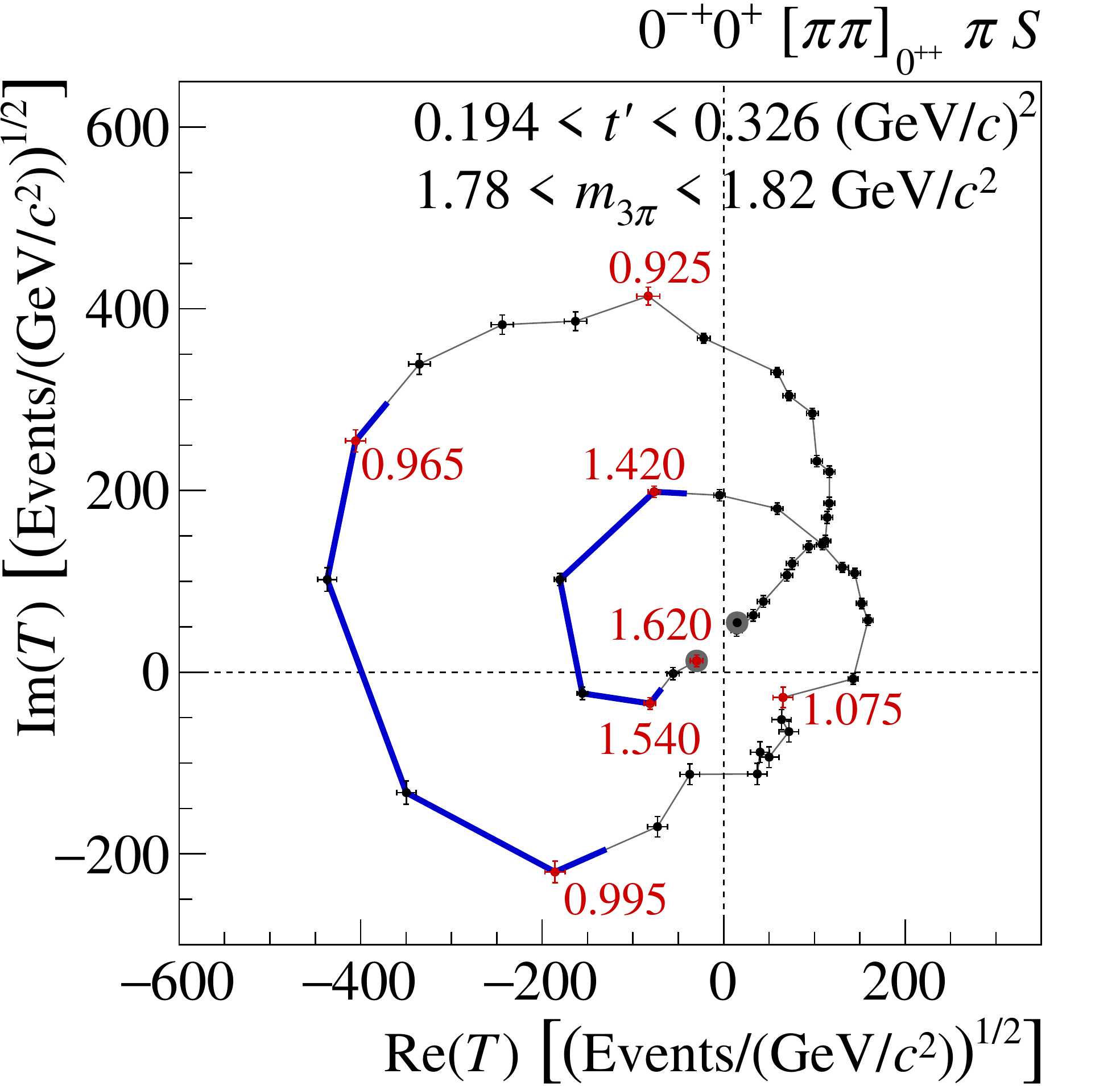}
\includegraphics[width=.33\columnwidth]{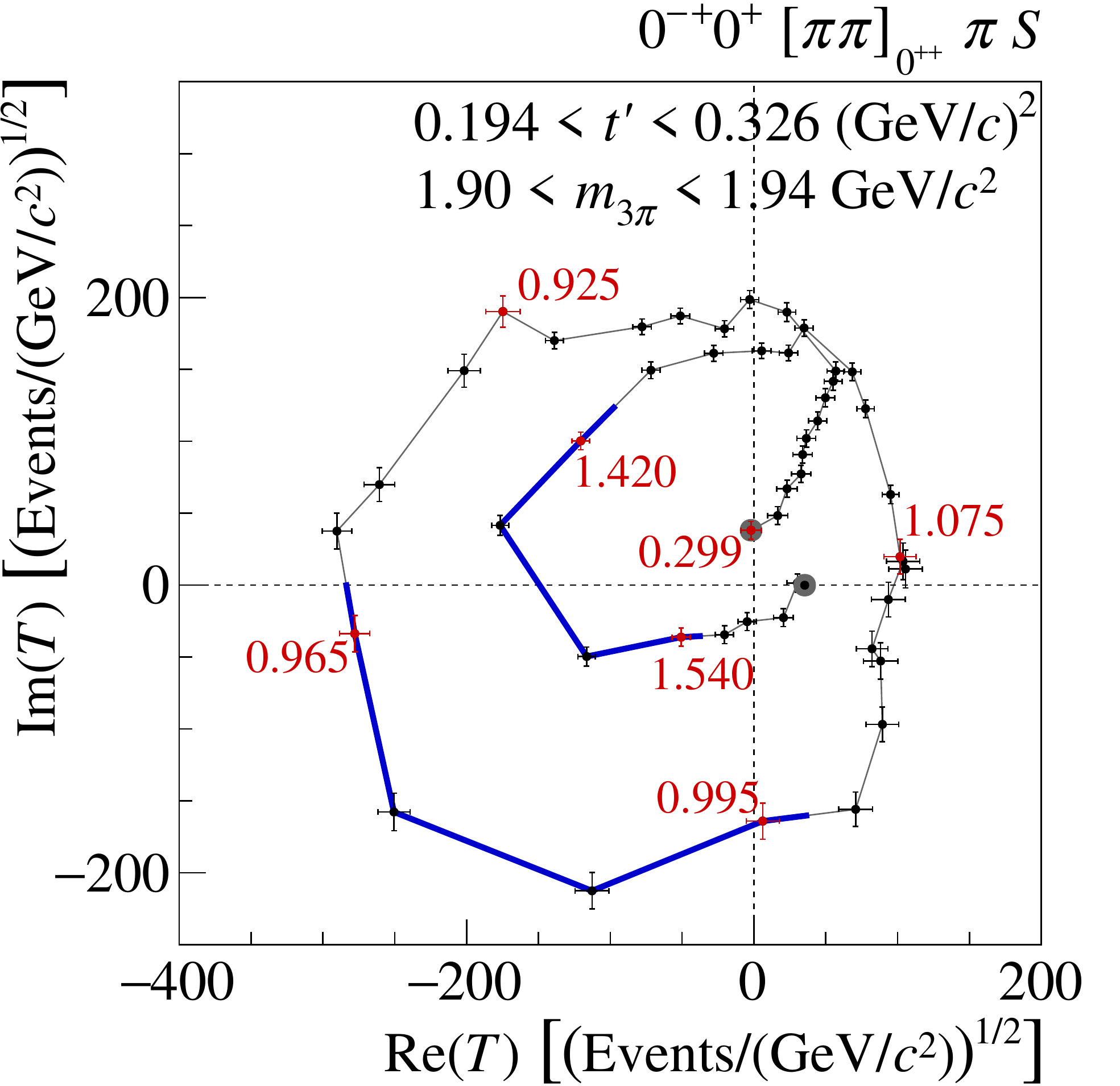}\\
\caption{Argand diagrams of the $0^{-+}0^+[\pi\pi]_{0^{++}}\pi S$ amplitude for three bins of $m_{3\pi}$ below, on, and above the $\pi(1800)$ resonance. The $m_{\pi^+\pi^-}$ regions corresponding to the $f_0(980)$ and the $f_0(1500)$ are highlighted in blue \cite{Adolph:2015tqa}.}
\label{fig::0mpArg}
\end{figure}

Fig. \ref{fig::0mpInt} and Fig. \ref{fig::0mpArg} show the intensity distributions and Argand diagrams as a function of $m_{\pi^+\pi^-}$ in narrow $m_{3\pi}$ bins around the $\pi(1800)$ resonance. Peaks and phase motions corresponding to the
$f_0(980)$ and the $f_0(1500)$ are visible. They are modulated by the intensity distribution and phase motion of the decay of $\pi(1800)$.


\subsection{$1^{++}0^+[\pi\pi]_{0^{++}}\pi\,P$ Wave}
\begin{figure}[h]
\begin{center}
\includegraphics[width=\widthTwo]{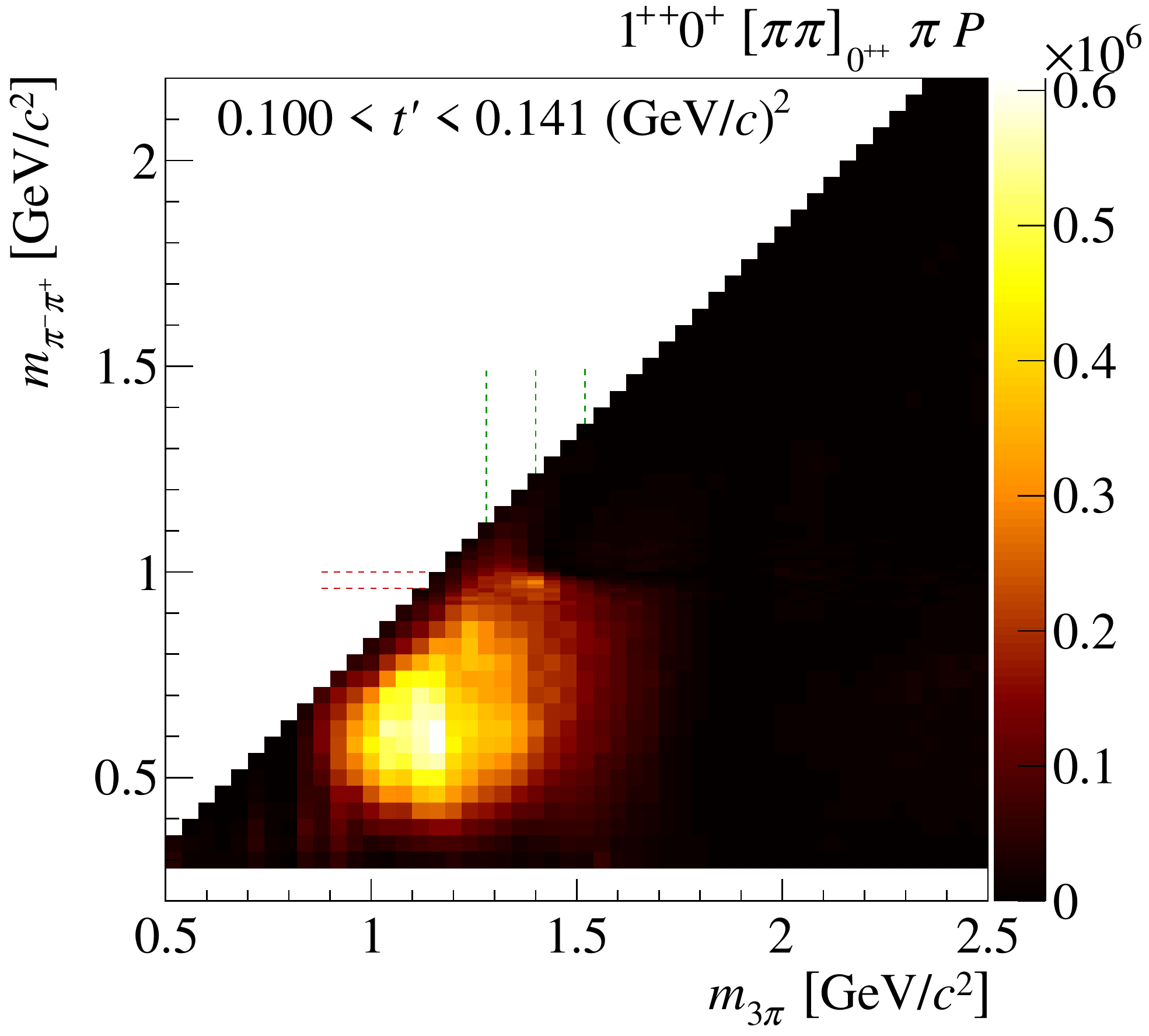}
\includegraphics[width=\widthTwo]{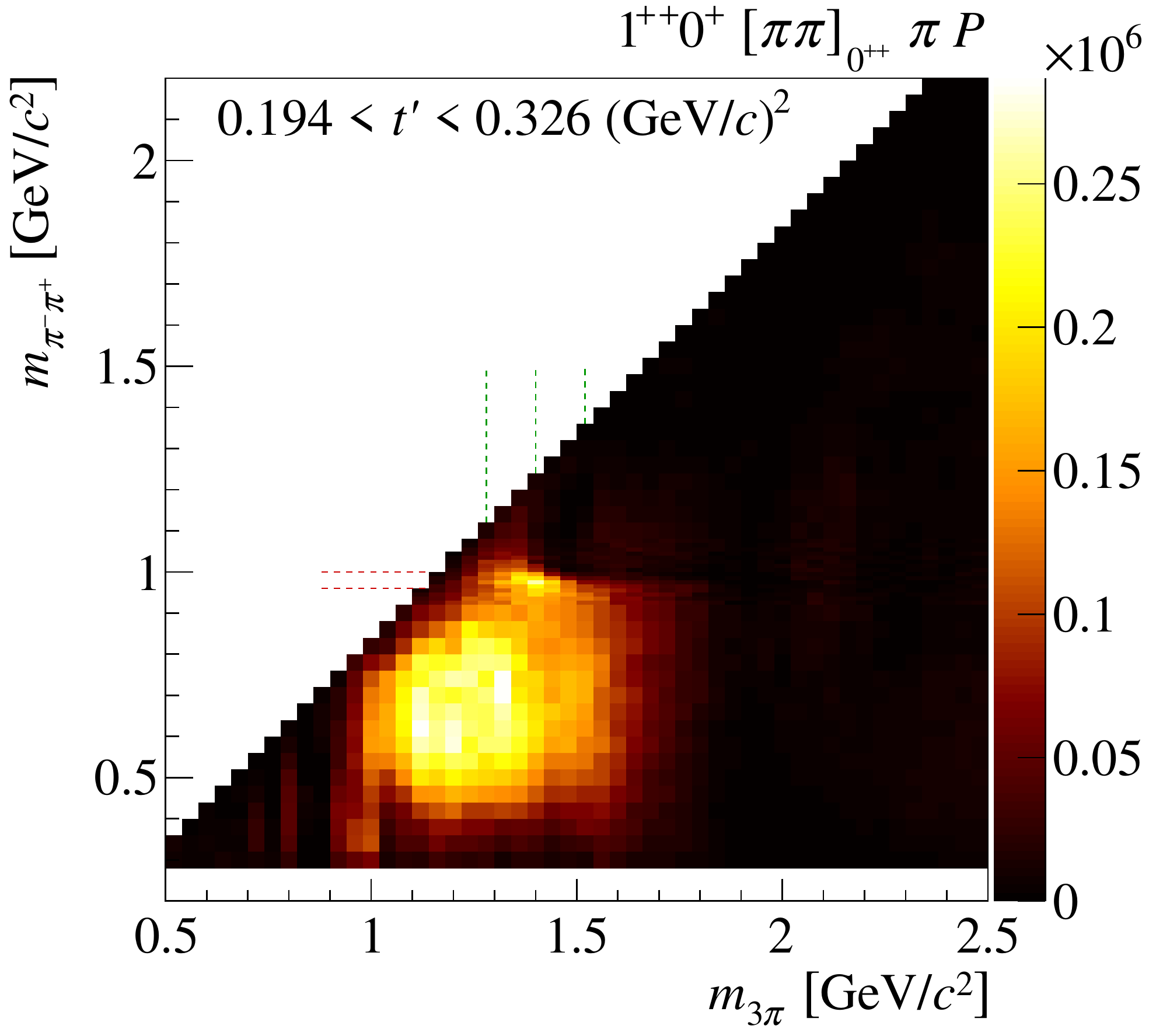}
\end{center}
\caption{Intensity distribution of the $1^{++}0^+[\pi\pi]_{0^{++}}\pi\,P$ wave as a function of $m_{3\pi}$ and $m_{\pipi}$ for two regions of $t^\prime$ \cite{Adolph:2015tqa}.}
\label{fig::1pp}
\end{figure}

\noindent The freed $1^{++}0^+[\pi\pi]_{0^{++}}\pi\,P$ wave is able to describe two waves of the conventional PWA, since the $1^{++}0^+$ $f_0(1500)\pi P$ wave was not included in the conventional analysis.
The two-dimensional intensity distribution is shown in Fig. \ref{fig::1pp} for two $t^\prime$ bins. It features a dominant broad structure at low $2\pi$ and $3\pi$ masses, which moves with $t^\prime$, 
indicating a predominantly non-resonant origin. In addition, a narrow peak at $m_{3\pi} \approx 1.4\,\GeV/c^2$ and a $m_{\pipi} \approx 0.98\,\GeV/c^2$ is visible. It corresponds to the recently discovered $a_1(1420)$ \cite{Adolph:2015pws}.
The observation of this peak in tha freed-isobar analysis proves that the $a_1(1420)$ signal is not an artifact of the parametrization of the scalar isobars in the conventional analysis \cite{Adolph:2015pws}.

\begin{figure}[h]
\includegraphics[width=.33\columnwidth]{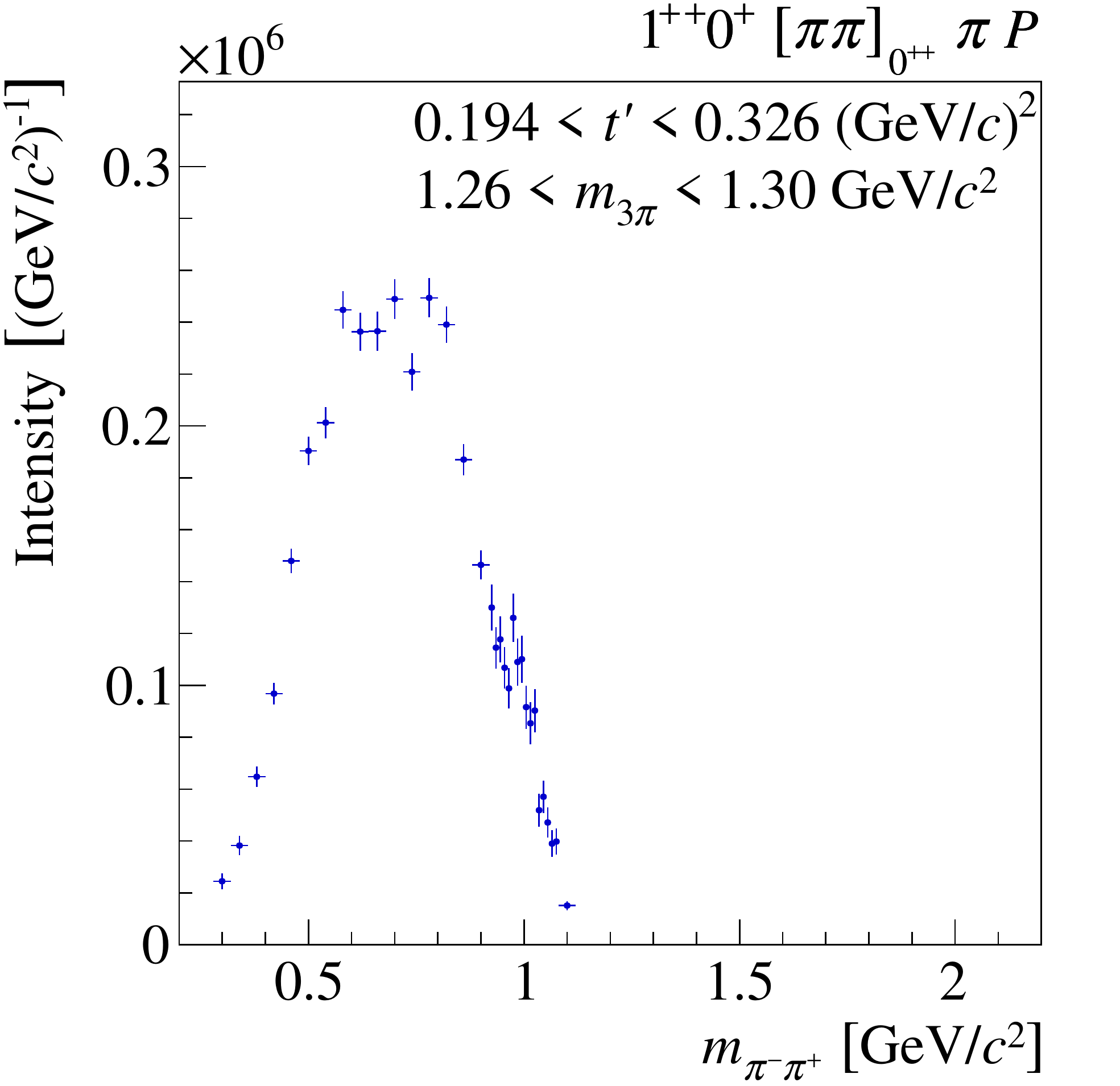}
\includegraphics[width=.33\columnwidth]{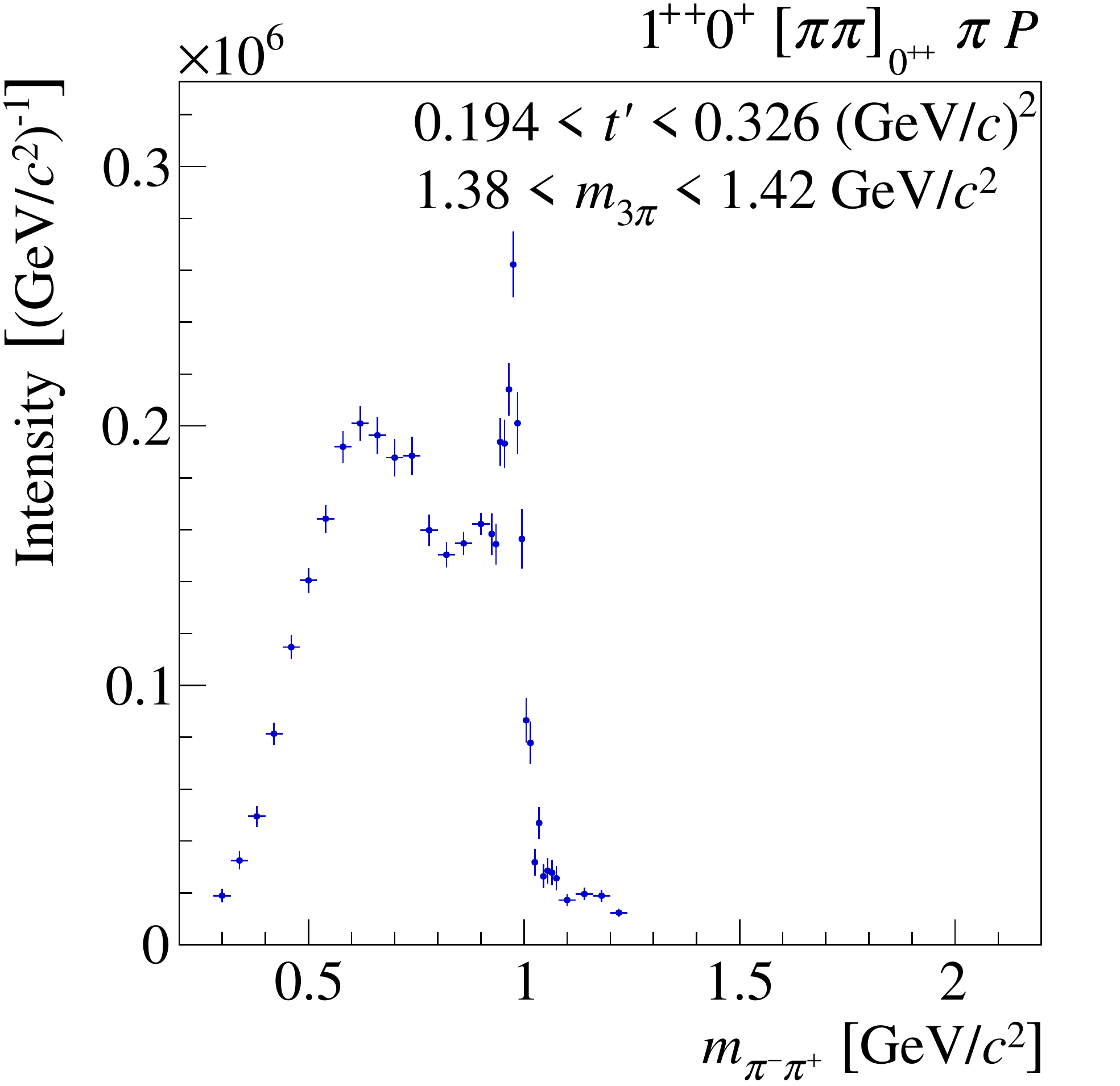}
\includegraphics[width=.33\columnwidth]{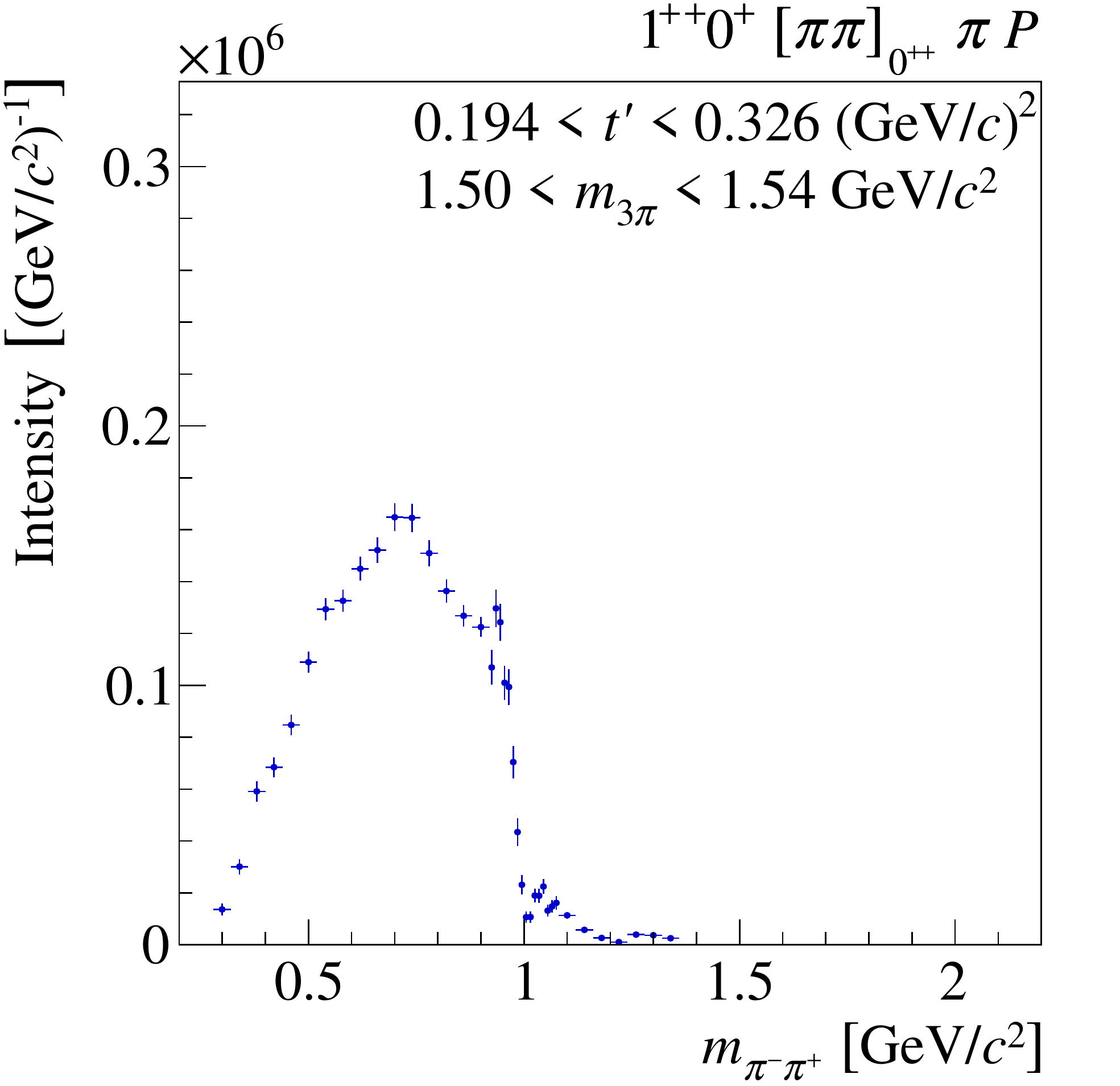}\\
\caption{$1^{++}0^+[\pi\pi]_{0^{++}}\pi P$ intensity distributions for three bins of $m_{3\pi}$ around the $a_1(1420)$ resonance \cite{Adolph:2015tqa}.}
\label{fig::1ppInt}
\end{figure}

\begin{figure}[h]
\begin{center}
\includegraphics[width=\widthTwo]{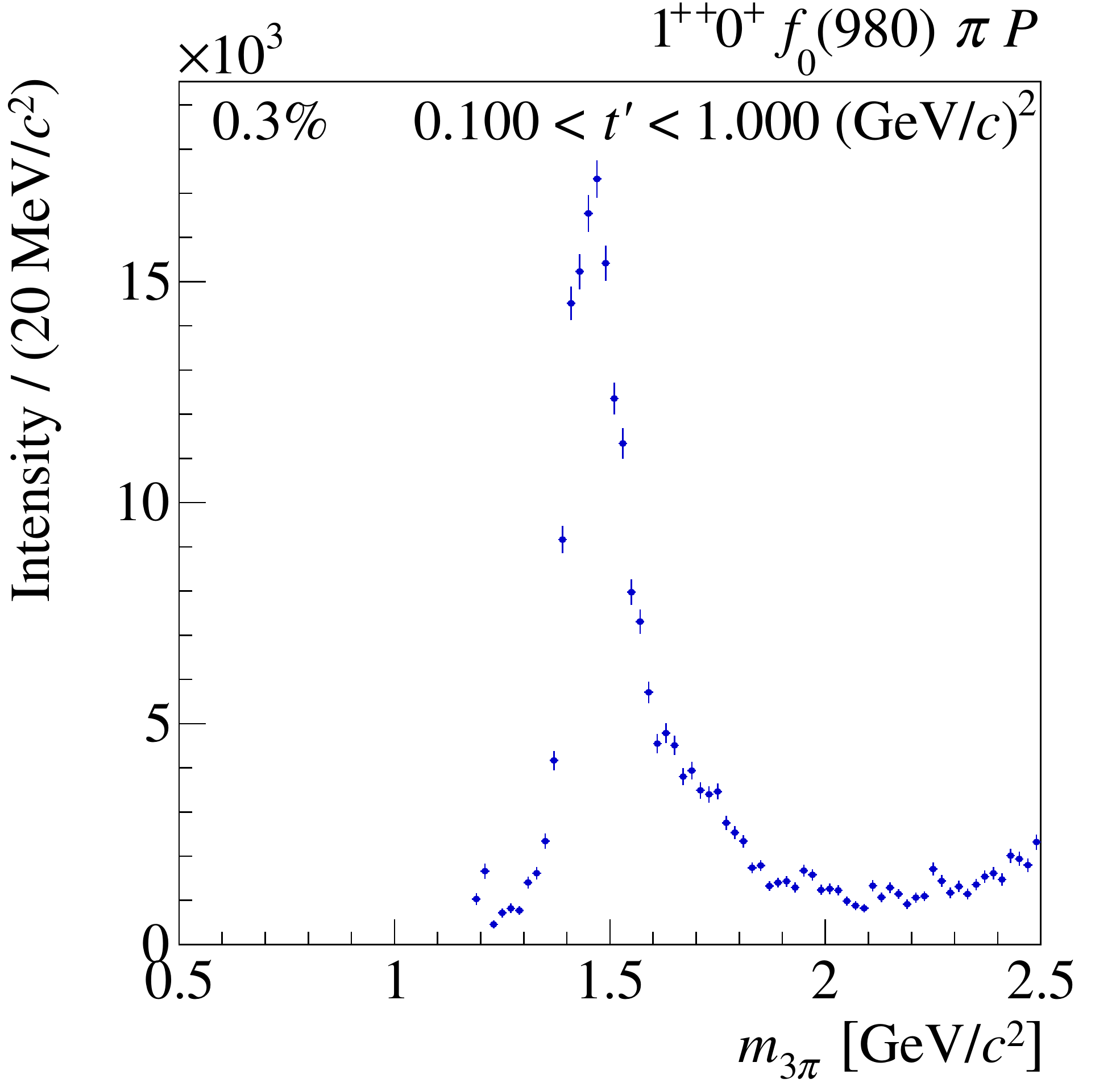}
\includegraphics[width=\widthTwo]{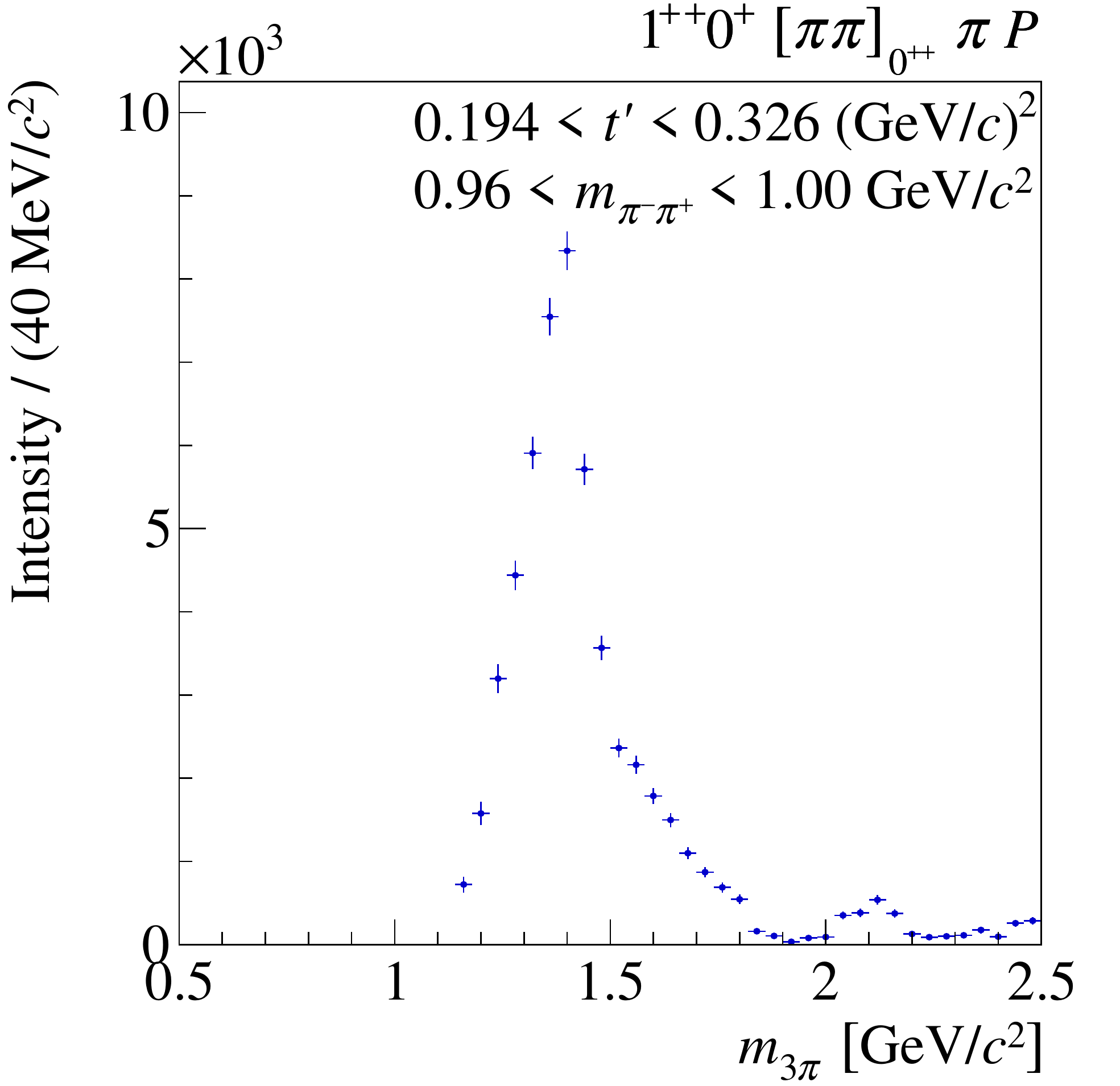}\\
\end{center}
\caption{$1^{++}0^+f_0(980)\pi P$ intensity from the conventional PWA (left) and intensity sum over the $m_{\pi^+\pi^-}$ bins in the $f_0(980)$ region of the $1^{++}0^+[\pi\pi]_{0^{++}}\pi P$ wave
from the freed-isobar analysis (right) \cite{Adolph:2015tqa}.}
\label{fig::1ppComp}
\end{figure}

Fig. \ref{fig::1ppInt} shows the $2\pi$ intensity distributions below, on, and above the $a_1(1420)$, which exhibits a strong correlation with the $f_0(980)$ peak. A comparison of the $f_0(980)$ mass region from the freed-isobar fit 
is in good agreement with the intensity of the $1^{++}0^+f_0(980)\pi P$ wave from the conventional PWA (See Fig. \ref{fig::1ppComp}).

\section{Conclusions}
\label{sec::conclusion}
We have introduced a novel PWA method for the process $\pi^-p\to\pi^-\pi^+\pi^-p$ using binned amplitudes to describe the $\pi^+\pi^-$ subsystems. This not only removes the model bias introduced by formerly fixed amplitudes used for the appearing isobars in 
the conventional PWA, but also allows us to study the $2\pi$ subsystems and their dependence on the $3\pi$ source system.

The large data set collected by the \textsc{Compass} spectrometer enables us to apply this novel method. 
In a first analysis, we free the $0^{++}$ isobar prametrizations for three waves with different $J^{PC}$ of the $3\pi$ parent system, namely $0^{-+}$, $1^{++}$, and $2^{-+}$. 
The analysis reproduces most of the expected structures, in particular a peak for the decay $a_1(1420)\to f_0(980)\pi$, which confirms the new signal observed with the conventional analysis not to be an artifact of the $f_0(980)$ parametrization.

In addition to resonances, broad structures are observed, that typically change their shape with $t^\prime$. They probably originate from non-resonant processes or from cross-talk with waves, that still employ fixed isobar amplitudes.

We are currently studying the latter effect by increasing the number of freed isobars. At the moment, we aim for a set of $11$ freed waves that would describe $75\%$ of the total intensity.
In these fits, we encounter several ambiguities in the fit and are currently working on techniques to resolve them.

\end{document}